# Prototypical π-π dimers re-examined by means of high-level CCSDT(Q) composite ab inito methods

Amir Karton[a,*] and Jan M. L. Martin[b]

[a]School of Molecular Sciences, The University of Western Australia, Perth, WA 6009, Australia.
[b]Department of Organic Chemistry, Weizmann Institute of Science, 76100 Reḥovot, Israel.

[*]Corresponding Author: amir.karton@uwa.edu.au (A.K.)

A B S T R A C T

The benzene•••ethene and parallel-displaced (PD) benzene•••benzene dimers are the most fundamental systems involving π-π stacking interactions. Several high-level ab initio investigations calculated the binding energies of these dimers at the CCSD(T)/CBS level of theory using various approaches such as reduced virtual orbital spaces and/or MP2-based basis set corrections. Here we obtain CCSDT(Q) binding energies using a Weizmann-3-type approach. In particular, we extrapolate the SCF, CCSD, and (T) components using large heavy-atom augmented Gaussian basis sets (namely, SCF/jul-cc-pV{5,6}Z, CCSD/jul-cc-pV{Q,5}Z, and (T)/jul-cc-pV{T,Q}Z). We consider post-CCSD(T) contributions up to CCSDT(Q), inner-shell, scalar-relativistic, and Born–Oppenheimer corrections. Overall, our best relativistic, all-electron CCSDT(Q) binding energies are $\Delta E_{e,all,rel}$ = 1.234 (benzene•••ethene) and 2.550 (benzene•••benzene PD), $\Delta H_0$ = 0.949 (benzene•••ethene) and 2.310 (benzene•••benzene PD), and $\Delta H_{298}$ = 0.130 (benzene•••ethene) and 1.461 (benzene•••benzene PD) kcal mol$^{-1}$. Important conclusions are reached regarding the basis set convergence of the SCF, CCSD, (T), and post-CCSD(T) components. Explicitly correlated calculations are used as a sanity check on the conventional binding energies. Overall, post-CCSD(T) contributions are destabilizing by 0.028 (benzene•••ethene) and 0.058 (benzene•••benzene) kcal mol$^{-1}$, thus they cannot be neglected if sub-chemical accuracy is sought (i.e., errors below 0.1 kcal mol$^{-1}$). CCSD(T)/aug-cc-pwCVTZ core-valence corrections increase the binding energies by 0.018 (benzene•••ethene) and 0.027 (benzene•••benzene PD) kcal mol$^{-1}$. Scalar-relativistic and diagonal Born–Oppenheimer corrections are negligibly small. We use our best CCSDT(Q) binding energies to evaluate the performance of MP2-based, CCSD-based, and lower-cost composite ab initio procedures for obtaining these challenging π-π stacking binding energies.

**Introduction**

Dispersion π-π interactions are ubiquitous in many biological and nano-chemical processes. For example, in (i) structures and dynamics of biomacromolecules such as DNA and proteins, (ii) biomolecular recognition and substrate binding, (iii) self-assembly in supramolecular systems, and (iv) solvation processes.[1,2,3,4,5] The binding energies in small π-stacked dimers generally vary between 1–5 kcal mol$^{-1}$. For example, for the seven dispersion-dominated π-π stacked dimers in the S66 database the interaction energies range between 1.4 (benzene•••ethene) and 5.6 (benzene•••uracil) kcal mol$^{-1}$.[6,7] The calculation of π-interactions in prototypical π-stacked dimers such as the benzene•••ethene and benzene•••benzene dimers to within chemical accuracy (arbitrarily defined as errors ≤ 0.1 kcal mol$^{-1}$ for weak noncovalent interactions) has been a long-standing goal of quantum chemistry. However, this has proved to be a challenging task due to the need to recover a significant portion of the correlation energy coupled with the exceedingly slow basis set convergence of the correlation component of the π-π interaction energy.[3,8,9,10,11,12,13,14,15,16,17,18]

It is generally accepted that coupled-cluster with singles, doubles, and quasi-perturbative triple excitations (CCSD(T)) recovers a sufficient portion of the correlation energy in order to achieve chemical accuracy for π-π interactions.[3] However, due to the steep computational scaling of the CCSD(T) method, it is a major challenge to approach the CCSD(T) one-particle basis set limit to within 0.1 kcal mol$^{-1}$. It is well established that a computationally efficient way to approach the CCSD(T)/CBS limit is to use the following MP2-based additivity scheme

$$E[\text{CCSD(T)/LBS}] \approx E[\text{CCSD(T)/SBS}] + E[\text{MP2/LBS}] - E[\text{MP2/SBS}] \quad (1)$$

where SBS and LBS indicate, respectively, small and large basis sets (or basis set extrapolations).[15] We denote this scheme as CCSD(T)[SBS]/MP2[LBS]. Sinnokrot and Sherrill calculated an interaction energy of 2.63 kcal mol$^{-1}$ from CCSD(T)[jul-cc-pVDZ]/MP2[aug-cc-pVQZ(no g)] calculations, where aug-cc-pVQZ(no g) indicates a truncated version of the aug-cc-pVQZ basis set basis in which g functions on carbon and f functions on hydrogen were omitted.[8] Hobza and co-workers[13] obtained interaction energies of 2.70 and 2.74 kcal mol$^{-1}$ using different extrapolation schemes from counterpoise (CP) corrected CCSD(T)[aug-cc-pV{T,Q}Z]/MP2[aug-cc-pV{Q,5}Z] energies. We note that the CCSD(T) calculations in this approach were performed in a truncated optimized virtual orbital space which significantly reduces the computational cost. Kesharwani et al.[18] combined CP-corrected MP2-F12/aug-cc-pV{T,Q}Z-F12 + [CCSD(F12*) – MP2-F12]/aug-cc-pVTZ-F12 + (T)/jul-cc-pV{D,T}Z and obtained an interaction energy of 2.69 kcal mol$^{-1}$. More recently, Alessandrini et al.[19] used a composite scheme which was developed specifically for noncovalent interactions and obtained an interaction energy of 2.77 kcal mol$^{-1}$. This scheme combines CCSD(T)[jun-cc-pVTZ]/MP2[jun-cc-pV{T,Q}Z] energies with a core-valence term calculated at the MP2 level. In addition, a number of high-level ab initio studies that did not use an MP2-based basis set correction were reported.[12,14] Janowski and Pulay obtained an interaction energy of 2.66 kcal mol$^{-1}$ from CP-corrected QCISD(T)/aug-cc-pV{T,Q}Z calculations,[12] whilst Sherrill et al.[14] obtained an almost identical interaction energy of 2.67 kcal mol$^{-1}$ at the CCSD(T)/jul-cc-pV{T,Q}Z level of theory. Nevertheless, it should be pointed out that both approaches (i.e., obtaining the CCSD(T)/CBS limit with or without an MP2-based basis set correction) are valid and give comparable results as long as sufficiently large basis sets are used. For example, Sherrill et al.[14] obtained an interaction energy of 2.71 kcal mol$^{-1}$ at the CCSD(T)[aug-cc-pV{D,T}Z]/MP2[aug-cc-pV{T,Q}Z] level (see also ref. 20 for a comparison of the two approaches for a diverse benchmark set of noncovalent interactions).

In the present work, we obtain the binding energies of the benzene•••ethene and benzene•••benzene π-π dimers using the Weizmann-$n$ (W$n$) composite ab initio theories.[21,22,23,24,25,26,27,28] In this approach the self-consistent field (SCF), CCSD correlation, and (T) correlation components are extrapolated individually to the CBS limit from the largest basis sets that are computationally feasible for each component. In particular, we were able to extrapolate the SCF, CCSD, and (T) components from the jul-cc-pV{5,6}Z, jul-cc-pV{Q,5}Z, and jul-cc-pV{T,Q}Z basis set pairs, respectively. We also explicitly consider post-CCSD(T) contributions up to CCSDT(Q), as well as secondary energetic contributions such as core-valence and diagonal Born–Oppenheimer corrections.

**Computational Details**

The SCF, MP2, and CCSD(T) calculations are carried out in conjunctions with the heavy-atom augmented jul-cc-pV$n$Z basis sets ($n$ = D, T, Q, 5, 6).[29,30,31] These basis sets are denoted by A'V$n$Z throughout the text. All basis set extrapolations use the $E(L) = E_\infty + A/L^\alpha$ two-point extrapolation formula (where $L$ is the highest angular momentum represented in the basis set) with $\alpha$ = 5 and 3, respectively for the SCF and correlation components.[32] Basis set extrapolations using the A'V$n$Z and A'V($n$+1)Z basis sets are denoted by A'V{$n,n+1$}Z.

Valence post-CCSD(T) contributions up to CCSDT(Q) are calculated in conjunction with the cc-pVDZ basis set as well as truncated versions of the cc-pVDZ and cc-pVTZ basis sets. Regular cc-pV$n$Z basis sets are denoted by V$n$Z throughout the text. The truncated basis sets are the VDZ basis set without the p functions on hydrogens (denoted by VDZ(no p on H)), the VDZ basis set without the p functions on hydrogen and d functions on carbon (denoted by VDZ(no d)), and a basis set which combines the sp part of the VTZ basis set with the d function from the VDZ basis set on carbon and the s part of the VTZ basis set with the p function from the VDZ basis set on hydrogen (denoted by VTZ(no f 1d)). We note that the fully iterative CCSDT/VDZ(no p on H) calculation for the benzene dimer involves 12.1×10$^9$ amplitudes and submicrohartree convergence of the CCSDT energy required 15 iterations, where each iteration ran for about 30 hours on a dual Intel Xeon machine with 40 cores and 1024 GB of RAM. For the benzene•••ethene dimer we were able to run the CCSDT(Q)/VDZ(no p on H) and CCSDT(Q)/VDZ calculations with diffuse sp functions added on carbons, these basis sets are denoted by jun-VDZ(no p on H) and jun-VDZ, respectively.

The core-valence (CV) contribution is calculated at the CCSD(T) level with the core-valence weighted correlation-consistent aug-cc-pwCVTZ basis set.[33] The scalar relativistic contribution, in the second-order Douglas–Kroll–Hess approximation,[34] is obtained from relativistic CCSD(T)/A'VTZ-DK calculations.[35] The diagonal Born–Oppenheimer corrections (DBOCs) are calculated at the HF/A'VTZ level of theory and a valence correlation contribution (ΔDBOC) is calculated at the CCSD/VDZ level of theory, i.e., ΔDBOC = DBOC(CCSD/VDZ) – DBOC(HF/VDZ).

All the above CCSD(T) calculations were carried out with the Molpro 2016.1 program suite.[36] All post-CCSD(T) calculations were carried out within the frozen-core approximation using the MRCC program suite.[37] Absolute energies throughout are normally converged to at least $10^{-8}$ Hartree in the CCSD(T) and lower-cost calculations and to $10^{-6}$ Hartree in the more expensive post-CCSD(T) calculations. All DBOC calculations were carried out with the CFOUR program suite.[38]

As a sanity check on the basis set convergence of the MP2 and CCSD components, explicitly correlated calculations were carried out at using Molpro 2020.1[39] running on the Faculty of Chemistry HPC cluster at the Weizmann Institute. For the MP2 component, MP2-F12 with the aug-cc-pVTZ-F12 and aug-cc-pVQZ-F12 basis sets[40] was carried out with a geminal exponent γ = 1.0 and the program defaults for the auxiliary basis sets, which are aug-cc-pVTZ/JK,[41] aug-cc-pVnZ/MP2FIT,[42] and cc-pVnZ-F12/OptRI.[43] Post-MP2 corrections were evaluated at the CCSD(F12*) level[44] with the cc-pVTZ-F12 basis set with γ = 1.0 (attempts to use aug-cc-pVTZ-F12 foundered on numerical problems). BSSE (basis set superposition error) was probed using the standard Boys-Bernardi counterpoise method[45] for both F12 and orbital-based calculations; for in-depth discussion of the basis set convergence of counterpoise corrections to noncovalent interaction energies, see Burns et al.[16] for orbital-based, and Brauer et al.[46,47] for explicitly correlated methods. The superiority of CCSD(F12*) over CCSD-F12b and especially CCSD-F12a was established at great length in Ref. 48.

All structures reported in the present work (dimers and corresponding monomers) are fully optimized. Accordingly, all the reported energies are binding energies (i.e., difference in energy between the optimized dimer and optimized isolated monomers) rather than interaction energies (i.e., energy difference between the fully or partly optimized dimer and isolated monomers in their geometry in the dimer). The geometries and vibrational frequencies were obtained for all structures at the DSD-PBEP86-D3BJ/Def2-QZVPPD level of theory.[49,50,51,52,53] The structures of all monomers as well as the benzene•••ethene and benzene•••benzene parallel-displaced (PD) and T-shaped tilted (TT) dimers were verified to have all real harmonic frequencies.[54] Zero-point vibrational energies (ZPVEs) and enthalpic corrections are calculated within the rigid rotor-harmonic oscillator (RRHO) approximation at the same level of theory. The harmonic ZPVEs were scaled as recommended in Ref. 55. All geometry optimizations and frequency calculations were performed using the Gaussian 16 program suite.[56]

Finally, we evaluate the performance of a range of spin-component-scaled CCSD and Møller–Plesset perturbation theory procedures[57] (SCS-MP2,[58] SCS(MI)-MP2,[59] SCSN-MP2,[60] SCS-MP2-vdW,[61] and S2-MP2,[62] SOS-MP2,[63] SCS-CCSD,[64] and SCS(MI)-CCSD).[65] In addition, we consider the MP2.5 (average of MP2 and MP3)[66] and MP3.5 (average of MP3 and MP4)[67] procedures, as well as a number of computationally economical composite ab initio procedures (namely, G4(MP2),[68] G4(MP2)-6X,[69] G4,[70] CBS-QB3,[71] CBS-APNO,[72] W1, and W2).[21,22]

## Results and Discussion

**DSD-PBEP86-D3BJ/Def2-QZVPPD geometries.** Before proceeding to a detailed discussion of the W3-type results for the benzene•••ethene and benzene•••benzene π-π dimers, it is instructive to examine the energetic consequences of using DSD-PBEP86-D3BJ/Def2-QZVPPD geometries rather than previously published geometries. For this purpose, we calculate the binding energies using the fully optimized DSD-PBEP86-D3BJ/Def2-QZVPPD geometries obtained in the present work and the original geometries from the S66 database for a number of dimers involving benzene. We note that the reference geometries from the S66 database are @1.0$r_e$.[6,7] We find that at the CCSD(T)[A'VTZ]/MP2[A'V{T,Q}Z] level the differences in binding energies between the two geometries are small but chemically significant. Namely, they are (in absolute value) 0.00 (benzene•••ethene), 0.01 (benzene•••water), 0.05 (benzene•••ethyne), and 0.09 (benzene•••benzene PD) kcal mol$^{-1}$.

**SCF component.** Table 1 gives an overview of the basis set convergence of the SCF component of the benzene•••ethene and benzene•••benzene π-π dimers. The SCF energy converges fairly quickly to the complete basis set (CBS) limit. The smallest basis set that results in deviations below the 0.1 kcal mol$^{-1}$ threshold is A'VQZ with deviations of 0.037 (benzene•••ethene) and 0.057 (benzene•••benzene) kcal mol$^{-1}$. Notably, the A'V{T,Q}Z extrapolation results in deviations below 0.01 kcal mol$^{-1}$ for both dimers (Table 1). Deviations of merely –0.004 (benzene•••ethene) and –0.008 (benzene•••benzene) kcal mol$^{-1}$ obtained for the A'V{Q,5}Z extrapolation, indicate that our A'V{5,6}Z reference values should be within ~1 wavenumber from the infinite one-particle basis set limit.

**Table 1.** Basis set convergence of the SCF component of the intermolecular binding energy in the benzene•••ethene (BzEt) and benzene•••benzene (Bz$_2$) π-π dimers (in kcal mol$^{-1}$).[a]

|            | BzEt   | Bz$_2$  |
|------------|--------|---------|
| A'VDZ      | 0.375  | 0.937   |
| A'VTZ      | 0.129  | 0.218   |
| A'VQZ      | 0.037  | 0.057   |
| A'V5Z      | 0.009  | 0.013   |
| A'V6Z      | 0.004  | 0.005   |
| A'V{D,T}Z  | 0.092  | 0.109   |
| A'V{T,Q}Z  | 0.008  | 0.007   |
| A'V{Q,5}Z  | –0.004 | –0.008  |
| A'V{5,6}Z  | –3.328 | –4.510  |

[a]The tabulated values are deviations from the A'V{5,6}Z reference values which are given in the bottom line (the negative sign of the A'V{5,6}Z reference values indicates a repulsive binding energy at the SCF level).

As a by-product of the MP2-F12 calculations, we obtained the CABS-corrected HF/aug-cc-pVQZ-F12 binding energies for (Bz$_2$) of –4.468 kcal/mol without, and –4.473 kcal mol$^{-1}$ with, counterpoise correction. The latter value is in as close an agreement with the best orbital-based estimate as we can reasonably expect.

**MP2 correlation component.** Table 2 gives an overview of the basis set convergence of the MP2 correlation component. For the benzene•••benzene dimer, we were able to obtain MP2 A'V{5,6}Z reference values, whilst for the benzene•••ethene dimer the A'V6Z calculations were not possible due to linear dependency problems. Thus, we use the A'V{Q,5}Z and A'V{5,6}Z reference values for the benzene•••ethene and benzene•••benzene dimers, respectively. Nevertheless, it should be noted that the difference between the A'V{Q,5}Z and A'V{5,6}Z extrapolations for the benzene•••benzene dimer is merely 0.010 kcal mol$^{-1}$ (Table 2). This result indicates that the difference between the A'V{Q,5}Z and A'V{5,6}Z extrapolations for the benzene•••ethene dimer are likely to be smaller than 0.01 kcal mol$^{-1}$ since it exhibits faster basis set convergence than the benzene dimer.

**Table 2.** Basis set convergence of the MP2 correlation component of the intermolecular binding energy in the benzene•••ethene (BzEt) and benzene•••benzene (Bz$_2$) π-π dimers (in kcal mol$^{-1}$).[a]

|         | BzEt  | Bz$_2$ |
|---------|-------|--------|
| A'VDZ   | 0.463 | 1.265  |
| A'VTZ   | 0.348 | 0.809  |
| A'VQZ   | 0.151 | 0.325  |
| A'V5Z   | 0.077 | 0.171  |
| A'V6Z   | N/A   | 0.099  |
| A'V{D,T}Z | 0.299 | 0.617 |
| A'V{T,Q}Z | 0.007 | –0.029 |
| A'V{Q,5}Z | 5.712 | 0.010 |
| A'V{5,6}Z | N/A   | 9.293  |

[a]The tabulated values are deviations from the A'V{Q,5}Z reference values for the benzene•••ethene dimer and A'V{5,6}Z reference values for the benzene•••benzene dimer. The reference values are given in the bottom line.

As expected, the MP2 correlation energy converges much slower than the SCF energy. For the benzene•••ethene dimer, the smallest basis set that results in a deviation below the chemical accuracy threshold is A'V5Z with a deviation of 0.077 kcal mol$^{-1}$. For the benzene•••benzene dimer, even the A'V6Z basis set barely achieves chemical accuracy with a deviation of 0.099 kcal mol$^{-1}$ from the A'V{5,6}Z basis set limit reference value. We note, however, that for both dimers the A'V{T,Q}Z extrapolation results in deviations that are below the chemical accuracy threshold, namely 0.007 (benzene•••ethene) and –0.029 (benzene•••benzene) kcal mol$^{-1}$. The deviation of merely 0.010 kcal mol$^{-1}$ between the A'V{Q,5}Z and A'V{5,6}Z extrapolations for the benzene•••benzene dimer indicates that our A'V{5,6}Z basis set limit reference values should be a few wavenumbers from the infinite one-particle basis set limit.

At the MP2-F12 level for the benzene dimer, we obtain raw(counterpoise) 9.360(9.297) kcal mol$^{-1}$ with the aug-cc-pVTZ-F12 basis set, and 9.324(9.309) kcal mol$^{-1}$ with the aug-cc-pVQZ-F12 basis set. Using the extrapolation exponent 4.6324 obtained in Ref. 18 for this basis set pair, the MP2-F12/aug-cc-pV{T,Q}Z-F12 extrapolation to the basis set limit yields 9.311(9.314) kcal mol$^{-1}$ — note that raw and counterpoise-corrected CBS limits should be equivalent, which is clearly as well satisfied as we can expect here. It was previously noted[27,46,47] that one advantage of F12 methods is that they dramatically reduce BSSE compared to orbital approaches in comparable-sized basis sets, by an order of magnitude or more.

The best orbital-based MP2 correlation contribution is within 0.02 kcal/mol, making the choice between orbital and F12 best estimates somewhat arbitrary: we have chosen to retain the orbital-based value, but now buttressed by its close correspondence to a result obtained by another (i.e., F12) approach.

**CCSD correlation component.** For both dimers our best CCSD correlation reference values are extrapolated from the A'V{Q,5}Z basis set pair. Table 3 lists the deviations from theses reference values for the individual basis sets and basis set extrapolations. As expected, the CCSD correlation component converges exceedingly slowly to the CBS limit. As is the case for the MP2 correlation energy for the benzene•••ethene dimer, the smallest basis set that results in a deviation below the chemical accuracy threshold is A'V5Z with a deviation of +0.067 kcal mol$^{-1}$. The A'V{T,Q}Z extrapolation performs somewhat better with a deviation of –0.051 kcal mol$^{-1}$.

**Table 3.** Basis set convergence of the CCSD correlation component of the intermolecular binding energy in the benzene•••ethene (BzEt) and benzene•••benzene (Bz$_2$) π-π dimers (in kcal mol$^{-1}$).[a]

|   | BzEt | Bz$_2$ |
|---|---|---|
| A'VDZ | 0.629 | 1.459 |
| A'VTZ | 0.382 | 0.833 |
| A'VQZ | 0.131 | 0.283 |
| A'V5Z | 0.067 | 0.145 |
| A'V{D,T}Z | 0.278 | 0.569 |
| A'V{T,Q}Z | –0.051 | –0.118 |
| A'V{Q,5}Z | 3.819 | 5.801[b] |
| A'V{Q,5}Z[c] | 3.779 | 5.797[d] |

[a]The tabulated values are deviations from the A'V{Q,5}Z reference values which are given in the bottom line. [b]An estimated CCSD/A'V{5,6}Z value using an MP2-based basis set correction results in a similar binding energy of 5.791 kcal mol$^{-1}$ (see text). [c]BSSE corrected (see text). [d]An estimated CCSD/A'V{5,6}Z value using an MP2-based basis set correction results in a similar binding energy of 5.787 kcal mol$^{-1}$ (see text).

For any given basis set or basis set extrapolation the deviation from the A'V{Q,5}Z reference values is about twice as large for the benzene•••benzene dimer than the benzene•••ethene dimer. Thus, none of the basis sets or basis set extrapolations result in deviations below the chemical accuracy threshold for the benzene•••benzene dimer. The A'V5Z basis set and A'V{T,Q}Z extrapolation are approaching chemical accuracy with deviations of 0.145 and –0.118 kcal mol$^{-1}$, respectively.

It should be pointed out that since the deviation between the A'V{T,Q}Z and A'V{Q,5}Z extrapolations for the benzene•••benzene dimer is just above the chemical accuracy threshold (–0.118 kcal mol$^{-1}$) we expect that our A'V{Q,5}Z reference value should be less than 0.1 kcal mol$^{-1}$ away from the infinite one-particle basis set limit. In an attempt to improve the CCSD/A'V{Q,5}Z reference value we can use an MP2-based basis set correction extrapolated from the A'V{5,6}Z basis set pair:

$$\Delta E[\text{CCSD/A'V}\{5,6\}Z] \approx \Delta E[\text{CCSD/A'V}\{Q,5\}Z] + \Delta E[\text{MP2/A'V}\{5,6\}Z] - \Delta E[\text{MP2/A'V}\{Q,5\}Z] \quad (2)$$

This results in an estimated CCSD/A'V{5,6}Z value of 5.791 kcal mol$^{-1}$, the MP2-based basis set correction in eq. (2) being merely –0.010 kcal mol$^{-1}$ (Table 2).

Finally, a note is due on the basis set superposition error (BSSE) for the CCSD correlation component. For the benzene•••ethene dimer we obtain extrapolated BSSE corrections of –0.082 (A'V{T,Q}Z) and –0.040 (A'V{Q,5}Z) kcal mol$^{-1}$. For the benzene•••benzene dimer we obtain extrapolated BSSE corrections of –0.179 (A'V{T,Q}Z) and –0.004 (A'V{Q,5}Z) kcal mol$^{-1}$. Inclusion of the A'V{Q,5}Z BSSE corrections leads to our best CCSD/CBS correlation components of 3.779 (benzene•••ethene) and 5.787 (benzene•••benzene) kcal mol$^{-1}$.

For benzene dimer, the best values correspond to a [CCSD–MP2]/A'V{Q,5}Z contribution of –3.502 kcal/mol. At the CCSD(F12*)/cc-pVTZ-F12 level, we obtain –3.472 kcal/mol without,

and –3.454 with counterpoise correction. The residual counterpoise uncertainty is of a similar order of magnitude as the gap with the conventional value, hence we have elected to retain the latter.

**(T) correlation component.** Table 4 gives an overview of the basis set convergence of the (T) correlation component relative to our best A'V{T,Q}Z reference values. The (T) component converges remarkably fast to the CBS limit and even the A'VDZ basis set results in sub-chemical accuracy with deviations of –0.020 (benzene•••ethene) and 0.091 (benzene•••benzene) kcal mol$^{-1}$. These deviations are reduced by a factor of three with the A'VTZ basis set. Overall, the A'V{D,T}Z and A'V{T,Q}Z basis set extrapolations result in practically indistinguishable binding energies, indicating that the latter are fully converged to the infinite one-particle basis set limit. The extrapolated A'V{T,Q}Z BSSE corrections are 0.001 (benzene•••ethene) and –0.008 (benzene•••benzene) kcal mol$^{-1}$. Adding these BSSE corrections leads to our best (T)/CBS correlation components of 0.793 (benzene•••ethene) and 1.303 (benzene•••benzene) kcal mol$^{-1}$. Overall, our best valence CCSD(T)/CBS binding energies are 1.244 (benzene•••ethene) and 2.580 (benzene•••benzene) kcal mol$^{-1}$.

**Table 4.** Basis set convergence of the (T) correlation component of the intermolecular binding energy in the benzene•••ethene (BzEt) and benzene•••benzene (Bz$_2$) π-π dimers (in kcal mol$^{-1}$).[a]

|            | BzEt   | Bz$_2$ |
|------------|--------|--------|
| A'VDZ      | –0.020 | 0.091  |
| A'VTZ      | –0.007 | 0.029  |
| A'VQZ      | –0.003 | 0.012  |
| A'V{D,T}Z  | –0.002 | 0.003  |
| A'V{T,Q}Z  | 0.792  | 1.310  |
| A'V{T,Q}Z[b] | 0.793 | 1.303  |

[a]The tabulated values are deviations from the A'V{T,Q}Z reference values which are given in the bottom line. [b]BSSE corrected (see text).

**Post-CCSD(T) contributions.** A number of high-level studies have considered the effect of post-CCSD(T) contributions on noncovalent interactions.[73,74,75,76,77,78,79,80,81,82,83,84] However, for reasons of computational cost, these studies considered fairly small dimers with up to 4 non-hydrogen atoms. Hobza and co-workers considered the π-π ethene•••ethyne, ethene•••ethyne, and ethyne•••ethyne dimers.[75,82] In Ref. 75 the CCSDT(Q)–CCSD(T) component was calculated in conjunction with a modified version of the 6-31G** basis set optimized for noncovalent interactions,[85] and it was found to be –0.027, –0.025, –0.031 kcal mol$^{-1}$ for the three dimers, respectively. In ref. 82 the CCSDT(Q)–CCSD(T) component was calculated in conjunction with the larger aug-cc-pVDZ basis set, however, overall similar results were obtained, namely contributions of –0.023, –0.024, –0.024 kcal mol$^{-1}$ were obtained for the three dimers, respectively. These results indicate that there may be some degree of error cancellation between basis set incompleteness of the T–(T) and (Q) components, as was also found by Patkowski and co-workers for some weakly bound polar and nonpolar complexes.[77] Hobza and co-workers[13] calculated post-CCSD(T) contributions to the binding energies of the PD and T-shaped benzene dimers using the computationally economical CCSD(TQ$_f$) methods which includes both the connected triple and quadruple excitations in a noniterative manner.[86] At the CCSD(TQ$_f$)/6-31G*(0.25) they estimated that post-CCSD(T) contributions are destabilizing by about 0.04 (PD) and 0.02 kcal mol$^{-1}$ (T and TT) dimers. In the present work, at great computational cost, we consider post-CCSD(T) contributions for the benzene•••ethene and benzene•••benzene π-stacked dimers at the CCSDT(Q) level with larger regular and partially augmented basis sets.

We begin by examining the post-CCSD(T) contributions to the binding energies of the benzene•••ethene dimer listed in Table 5. We were able to calculate the CCSDT–CCSD(T) (T–(T)) difference with basis sets of up to VTZ(no f 1d) quality and obtain the following contributions: –0.037 (VDZ(no d)), –0.080 (VDZ(no p on H)), –0.081 (VDZ), and –0.091 (VTZ(no f 1d)) kcal mol$^{-1}$. Thus, omitting the d functions from carbon atoms in the VDZ(no d) basis set results in a significant underestimation of the T–(T)/VTZ(no f 1d) component by over 50%. However, both the VDZ(no p on H) and VDZ basis sets results in a T–(T) contribution of –0.08 kcal mol$^{-1}$ which underestimates the T–(T)/VTZ(no f 1d) component by merely 0.01 kcal mol$^{-1}$ (Table 5).

**Table 5.** Secondary energetic contributions to the intermolecular binding energy in the benzene•••ethene (BzEt) and benzene•••benzene (Bz$_2$) π-π dimers (in kcal mol$^{-1}$).

| Component | | BzEt | Bz$_2$ |
|---|---|---|---|
| T–(T) | VDZ(no d) | –0.037 | –0.100 |
| T–(T) | VDZ(no p on H) | –0.080 | –0.186 |
| T–(T) | jun-VDZ(no p on H) | –0.094 | N/A |
| T–(T) | VDZ | –0.081 | –0.194[a] |
| T–(T) | jun-VDZ | –0.092 | N/A |
| T–(T) | VTZ(no f 1d) | –0.091 | –0.246[a] |
| (Q) | VDZ(no d) | 0.044 | 0.080 |
| (Q) | VDZ(no p on H) | 0.065 | 0.200[b] |
| (Q) | jun-VDZ(no p on H) | 0.081 | N/A |
| (Q) | VDZ | 0.063 | 0.187[b] |
| (Q) | jun-VDZ | 0.078 | N/A |
| Best post-CCSD(T) | | –0.028[c] | –0.058[c] |
| CV | APWCVTZ | 0.018 | 0.027 |
| Scalar rel. | A'VTZ-DK | 0.002 | 0.003 |
| DBOC | HF/A'VTZ | –0.003 | –0.003 |
| DBOC[d] | ΔCCSD | 0.001 | 0.002 |
| Best DBOC[e] | | –0.002 | –0.001 |
| Best secondary cont.[f] | | –0.010 | –0.030 |

[a]Obtained from reaction (2) where the reaction energy is calculated with the VDZ(no p on H) basis set (see main text). [b]Obtained from reaction (2) where the reaction energy is calculated with the VDZ(no d) basis set (see main text). [c]T–(T)/VTZ(no f 1d) + (Q)/VDZ. [d]ΔCCSD = DBOC(CCSD/VDZ) – DBOC(HF/VDZ). [e]Sum of the HF/A'VTZ and ΔCCSD contributions. [f]Sum of best post-CCSD(T), CV, scalar rel., and DBOC contributions.

In the preceding paragraph we have seen that the VDZ(no p on H) and VDZ basis sets provide a very good approximation to the T–(T)/VTZ(no f 1d) component. However, an important question is what would be the effect of adding diffuse functions on the carbon atoms? To answer this question, we were able to add a set of diffuse sp functions on the carbon atoms in the VDZ(no p on H) and VDZ basis sets (denoted by jun-VDZ(no p on H) and jun-VDZ, respectively). In both cases, addition of the set of sp diffuse functions affects the T–(T) component by a fairly small amount of ~0.01 kcal mol$^{-1}$, i.e., an order of magnitude below our target accuracy (see Table 5). Ideally, we should be calculating the T–(T) contributions with heavy-atom augmented basis sets (i.e., including diffuse d functions on carbon), however, these calculations proved beyond our computational resources. For example, the fully iterative CCSDT calculations involve $3.0\times10^9$ (jun-VDZ(no p on H)) and $5.4\times10^9$ (jun-VDZ) amplitudes and the quasiperturbative CCSDT(Q) calculations involve $1.7\times10^{12}$ (jun-VDZ(no p on H)) and $3.8\times10^{12}$ (jun-VDZ) determinants in the (Q) step.

For the CCSDT(Q)–CCSDT ((Q)) difference in the benzene•••ethene dimer we obtain the following contributions: 0.044 (VDZ(no d)), 0.065 (VDZ(no p on H)), and 0.063 (VDZ) kcal mol$^{-1}$. Thus, similarly to the T–(T) component, the VDZ(no d) basis set results in a significant underestimation of the (Q)/VDZ component by ~30%, but omitting the p functions from hydrogens has practically no effect on the (Q) component. Adding a set of diffuse sp functions on carbons atoms affects the (Q) component to a slightly larger extent than the T–(T) component. Namely, the (Q) component is increased by 0.016 (VDZ(no p on H) → jun-VDZ(no p on H)) and 0.015 (VDZ → jun-VDZ) kcal mol$^{-1}$. However, these changes are still well below our target chemical accuracy of 0.1 kcal mol$^{-1}$.

It is important to note that similarly to atomization energies,[24,25,27,28] the T–(T) contribution is repulsive (i.e., it reduces the binding energy) whilst the (Q) contribution is attractive (i.e., it increases the binding energy). Thus, overall post-CCSD(T) contributions involve a significant degree of cancelation between the T–(T) and (Q) components. In addition, since the addition of diffuse functions makes the T–(T) more negative and the (Q) component more positive, their effect on the overall post-CCSD(T) contributions is even smaller than on the individual T–(T) and (Q) components. Overall, we obtain the following post-CCSD(T) contributions –0.015 (VDZ(no p on H)), –0.013 (jun-VDZ(no p on H)), –0.019 (VDZ), –0.014 (jun-VDZ) kcal mol$^{-1}$. Thus, the effect of adding a set of diffuse functions on the overall post-CCSD(T) contributions amounts to merely 0.001 (VDZ(no p on H) → jun-VDZ(no p on H)) and 0.004 (VDZ → jun-VDZ) kcal mol$^{-1}$.

For the benzene•••benzene dimer we were only able to calculate the T–(T) contribution with the VDZ(no p on H) basis set, since the CCSDT/VDZ(no p on H) calculation involves $12.1 \times 10^9$ amplitudes. The T–(T)/VDZ(no p on H) contribution amounts to –0.186 kcal mol$^{-1}$. However, since we have the T–(T)/VTZ(no f 1d) contribution for the benzene•••ethene dimer we can obtain an approximated T–(T)/VTZ(no f 1d) contribution for the benzene•••benzene dimer from the following thermochemical cycle that aims to balance the π-π interactions on both sides of the reaction:

$$Bz_2 + 4\ Bz + 3\ Et_2 \rightarrow 6\ BzEt \qquad (3)$$

Calculating the T–(T) reaction energy in conjunction with the VDZ(no p on H) basis set and the T–(T) energies of all species involved apart from Bz$_2$ with the VDZ basis set results in an estimated T–(T)/VDZ contribution of –0.194 kcal mol$^{-1}$. It is perhaps not surprising that this value is less than 0.01 kcal mol$^{-1}$ away from the T–(T)/VDZ(no p on H) value. However, calculating the T–(T) energies of all species apart from Bz$_2$ with the VTZ(no f 1d) basis set results in a considerably larger estimated T–(T)/VTZ(no f 1d) contribution of –0.246 kcal mol$^{-1}$.

For the benzene•••benzene dimer we were able to calculate the (Q) contribution with the VDZ(no d) basis set. At this level of theory, we obtain a modest (Q) contribution of 0.080 kcal mol$^{-1}$. However, the results for the benzene•••ethene dimer indicate that this value could be severely underestimated. We can obtain a better estimation of the (Q) contribution via reaction (2). Calculating the reaction energy at the (Q)/VDZ(no d) level of theory and the (Q) energies of all species apart from Bz$_2$ with the VDZ(no p on H) basis set results in an estimated (Q)/VDZ(no P on H) contribution of 0.200 kcal mol$^{-1}$. Upgrading the basis set for the reaction species to the VDZ basis set results in a slightly smaller (Q)/VDZ contribution of 0.187 kcal mol$^{-1}$. Overall, the post-CCSD(T) contribution for the benzene•••benzene dimer at our best level of theory (T–(T)/VTZ(no f 1d) + (Q)/VDZ) amounts to –0.058 kcal mol$^{-1}$.

**Secondary energetic contributions.** Table 5 also lists the core-valence, scalar relativistic, and diagonal Born–Oppenheimer corrections. The CCSD(T)/aug-cc-pwCVTZ core-valence corrections increase the binding energies by 0.018 (benzene•••ethene) and 0.027 (benzene•••benzene) kcal mol$^{-1}$. The scalar relativistic contributions to the binding energies are, as expected, practically nil for both dimers. Similarly, our best DBOC corrections are negligibly small. Overall, the post-CCSD(T), core-valence, scalar relativistic, and diagonal Born–Oppenheimer corrections amount to –0.010 (benzene•••ethene) and –0.030 (benzene•••benzene) kcal mol$^{-1}$.

**All-electron, relativistic, DBOC-inclusive CCSDT(Q) binding energies.** Table 6 summarizes our best CCSDT(Q) binding energies on the $\Delta E_e$, $\Delta H_0$, and $\Delta H_{298}$ potential energy surfaces. The valence CCSDT(Q) binding energies are $\Delta E_{e,\text{val}}$ = 1.216 (benzene•••ethene) and 2.522 (benzene•••benzene) kcal mol$^{-1}$. Inclusion of the secondary energetic contributions (CV, scalar rel., and DBOC) increases the binding energies by 0.02–0.03 kcal mol$^{-1}$ to $\Delta E_{e,\text{all,rel}}$ = 1.234 (benzene•••ethene) and 2.550 (benzene•••benzene) kcal mol$^{-1}$.

**Table 6.** Best theoretical all-electron, relativistic, CCSDT(Q) binding energies for the benzene•••ethene (BzEt) and benzene•••benzene (Bz$_2$) π-π dimers on the electronic ($\Delta E_e$), enthalpic at 0 K ($\Delta H_0$), and enthalpic at 298 K ($\Delta H_{298}$) potential energy surfaces (in kcal mol$^{-1}$).

|  | BzEt | Bz$_2$ |
|---|---|---|
| $\Delta E_{e,\text{val}}$[a] | 1.216 | 2.522 |
| $\Delta E_{e,\text{all}}$[b] | 1.234 | 2.548 |
| $\Delta E_{e,\text{all,rel}}$[c] | 1.234 | 2.550 |
| ZPVE[d] | –0.286 | –0.240 |
| $\Delta H_0$[e] | 0.949 | 2.310 |
| $\Delta H_{298}$[f] | 0.130 | 1.461 |

[a]Best valence CCSDT(Q) value on the electronic surface. [b]Best all-electron CCSDT(Q) value on the electronic surface. [c]Best all-electron, relativistic, DBOC-inclusive CCSDT(Q) value on the electronic surface. [d]Scaled DSD-PBEP86-D3BJ/Def2-QZVPPD harmonic ZPVEs. [e]Best all-electron, relativistic, DBOC-inclusive, ZPVE-inclusive CCSDT(Q) value on the enthalpic surface at 0 K. [f]Best all-electron, relativistic, DBOC-inclusive, ZPVE-inclusive CCSDT(Q) value on the enthalpic surface at 298 K.

Scaling the DSD-PBEP86-D3BJ/Def2-QZVPPD harmonic ZPVE corrections by 0.9827 as recommended in ref. 55 results in ZPVE corrections of –0.286 (benzene•••ethene) and –0.240 (benzene•••benzene) kcal mol$^{-1}$. This results in relativistic, all-electron CCSDT(Q) binding energies at 0 K of $\Delta H_0$ = 0.949 (benzene•••ethene) and 2.310 (benzene•••benzene) kcal mol$^{-1}$. There are two experimental measurements of the binding energy of the benzene dimer.[87,88] Both measurements likely involve a distribution of the PD and TT local minima structures. Since the equilibrium geometry of the TT dimer has no symmetry, we were only able to obtain the binding energy for this system at the W1val level and obtain $\Delta E_e$ = 2.72 and $\Delta H_0$ = 2.38 kcal mol$^{-1}$. Overall, our highly accurate binding energies at 0 K ($\Delta H_0$) for the PD (2.310) and TT (2.38 kcal mol$^{-1}$) dimers are in excellent agreement with the experimental value of Grover et al.[87] of $\Delta H_0$ = 2.4 ± 0.4 kcal mol$^{-1}$, albeit the theoretical values are associated with smaller uncertainties. However, our best theoretical $\Delta H_0$ binding energies for the PD and TT dimers are higher by nearly 1 kcal mol$^{-1}$ than the experimental value of Krause et al.[88] $\Delta H_0$ = 1.6 ± 0.2 kcal mol$^{-1}$, and suggest that the experimental value should be significantly revised upwards.

Finally, we can also convert our binding energies at 0 K ($\Delta H_0$) to 298 K ($\Delta H_{298}$) using enthalpic temperature corrections ($H_{298}$–$H_0$) obtained within the rigid rotor harmonic approximation at the DSD-PBEP86-D3BJ/Def2-QZVPPD level of theory. This results in $\Delta H_{298}$ = 0.130 (benzene•••ethene) and 1.461 (benzene•••benzene) kcal mol$^{-1}$.

**Performance of computationally economical standard and composite ab initio methods.** Table 7 lists the deviations between our best CCSDT(Q) and MP$n$-based binding energies extrapolated from the A'V{D,T}Z and A'V{T,Q}Z basis sets, whilst Table S2 of the Supporting Information lists the results for the individual basis sets. With the exception of SOS-MP2 and MP3, all the MP$n$-based methods (MP2, SCS-MP2, SCS(MI)-MP2, SCSN-MP2, SCS-MP2-vdW, S2-MP2, and MP2.5) exhibit fairly smooth basis set convergence. Table 7 also lists the percentage errors and in the following discussion we define 10% as an arbitrary target accuracy. Inspection of the results in Table 7 reveals that SCS-MP2/A'V{T,Q}Z is the only MP$n$-based procedure that can achieve this target accuracy for both dimers. The SCS(MI)-MP2 and SCSN-MP2 procedures, which were developed for a range of noncovalent interactions, result in deteriorated performance relative to SCS-MP2. Specifically, in conjunction with the A'V{T,Q}Z basis set extrapolation, they result in percentage errors of up to about 20% for the benzene•••benzene dimer (Table 7). Of the higher order MP$n$ procedures (MP2.5, MP3, MP3.5, and MP4), MP3.5 achieves similar performance to SCS-MP2 at a significantly higher computational cost, whilst the other procedures do not significantly improve on the performance of the best performing MP2-based procedures. Both of the CCSD-based procedures (SCS-CCSD and SCS(MI)-CCSD) achieve good performance with percentage errors of ~5% (BzEt) and ~11% (Bz$_2$) in conjunction with the A'V{T,Q}Z basis set pair (Table 7). For comparison the CCSD/A'V{T,Q}Z level of theory results in percentage errors of over 50%.

**Table 7.** Performance of computationally economical standard and composite ab initio methods. The tabulated values are deviations from our best CCSDT(Q) binding energies for the benzene•••ethene (BzEt) and benzene•••benzene (Bz$_2$) π-π dimers (in kcal mol$^{-1}$) (percentage errors are listed in parenthesis).[a]

|  |  | BzEt | Bz$_2$ |
|---|---|---|---|
| MP2 | A'V{D,T}Z | 1.49 (123) | 2.79 (111) |
|  | A'V{T,Q}Z | 1.14 (94) | 2.16 (86) |
| SCS-MP2 | A'V{D,T}Z | 0.38 (31) | 0.95 (38) |
|  | A'V{T,Q}Z | –0.01 (1) | 0.24 (9) |
| SCS(MI)-MP2 | A'V{D,T}Z | 0.27 (22) | 0.78 (31) |
|  | A'V{T,Q}Z | 0.07 (6) | 0.47 (19) |
| SCSN-MP2 | A'V{D,T}Z | 0.19 (15) | 0.65 (26) |
|  | A'V{T,Q}Z | 0.09 (8) | 0.56 (22) |
| SCS-MP2-vdW | A'V{D,T}Z | 0.99 (82) | 1.96 (78) |
|  | A'V{T,Q}Z | 0.58 (48) | 1.21 (48) |
| SOS-MP2 | A'V{D,T}Z | –0.18 (15) | 0.03 (1) |
|  | A'V{T,Q}Z | –0.59 (49) | –0.72 (29) |
| S2-MP2 | A'V{D,T}Z | 1.33 (109) | 2.51 (100) |
|  | A'V{T,Q}Z | 0.94 (77) | 1.81 (72) |
| MP2.5 | A'V{D,T}Z | 0.73 (60) | 1.09 (43) |
|  | A'V{T,Q}Z | 0.37 (30) | 0.43 (17) |
| MP3 | A'V{D,T}Z | –0.03 (2) | –0.61 (24) |
|  | A'V{T,Q}Z | –0.41 (34) | –1.29 (51) |
| MP3.5 | A'V{D,T}Z | 0.36 (29) | 0.42 (17) |
|  | A'V{T,Q}Z | –0.04 (3) | –0.29 (11) |
| MP4 | A'V{D,T}Z | 0.74 (61) | 1.45 (58) |
|  | A'V{T,Q}Z | 0.32 (27) | 0.72 (29) |
| CCSD | A'V{D,T}Z | –0.42 (35) | –0.75 (30) |
|  | A'V{T,Q}Z | –0.81 (66) | –1.41 (56) |
| SCS-CCSD | A'V{D,T}Z | 0.40 (33) | 0.55 (22) |
|  | A'V{T,Q}Z | –0.07 (5) | –0.29 (12) |
| SCS(MI)-CCSD | A'V{D,T}Z | 0.37 (30) | 0.47 (19) |
|  | A'V{T,Q}Z | –0.06 (5) | –0.28 (11) |
| CCSD(T) | A'VDZ | 1.05 (86) | 2.57 (102) |
|  | A'VTZ | 0.57 (47) | 1.16 (46) |
|  | A'V{D,T}Z | 0.37 (30) | 0.57 (23) |
| CCSD(T)/CBS(MP2) | A'VDZ/A'V{D,T}Z[b] | 0.54 (44) | 0.90 (36) |
|  | A'VDZ/A'V{T,Q}Z[c] | 0.19 (15) | 0.26 (10) |
|  | A'VTZ/A'V{T,Q}Z[d] | 0.07 (5) | 0.03 (1) |
| G4(MP2) |  | 0.83 (69) | 1.80 (72) |
| G4(MP2)-6X |  | 0.70 (57) | 1.62 (64) |
| G4 |  | 0.70 (58) | 1.60 (63) |
| CBS-QB3 |  | 0.49 (41) | 1.56 (62) |
| CBS-APNO |  | 0.80 (66) | 1.94 (77) |
| W1 |  | 0.03 (2) | –0.02 (1) |
| W2 |  | 0.06 (5) | 0.07 (3) |

[a]The deviations for the standard and composite ab initio procedures are calculated relative to the $\Delta E_{e,\text{val}}$ and $\Delta E_{e,\text{all,rel}}$ CCSDT(Q) reference values in Table 6, respectively. [b]CCSD(T)[A'VDZ]/MP2[A'V{D,T}Z]. [c]CCSD(T)[A'VDZ]/MP2[A'V{T,Q}Z]. [d]CCSD(T)[A'VTZ]/MP2[A'V{T,Q}Z].

The CCSD(T)/A'VDZ level of theory results in percentage errors of 86–102%, and the CCSD(T)/A'VTZ level of theory still results in unacceptably high percentage errors of 46–47% (Table 7). Extrapolating the CCSD(T) binding energy from the A'V{D,T}Z basis set pair results in percentage errors that are still outside our target accuracy of 10%.

We also evaluate the performance of the cost-effective CCSD(T)/CBS(MP2) method, in which the CCSD(T)/CBS energy is estimated from the CCSD(T)/A'VDZ energy and an MP2-based basis-set-correction term ($\Delta$MP2 = MP2/A'V{$n,n+1$}Z – MP2/A'VDZ, where $n \geq$ D). As expected, extrapolating the $\Delta$MP2 term from the A'V{D,T}Z basis set pair results in fairly large percentage errors of over 30%. However, extrapolating the $\Delta$MP2 term from the A'V{T,Q}Z basis set pair results in a significant improvement in performance at a reasonable computational cost with percentage errors of 10–15%. We note that upgrading the basis set in the CCSD(T) calculations to A'VTZ results in absolute deviations below 0.1 kcal mol$^{-1}$ (in absolute value) and percentage errors ≤ 5%.

Let us move to the performance of the composite ab initio methods. The computationally economical G4- and CBS-type procedures use a CCSD(T)/6-31G(d) base energy in conjunction with MP$n$-based basis set correction terms. As might be expected, these composite procedures show relatively poor performance, which is largely attributed to the small basis sets used in the CCSD(T) and $\Delta$MP$n$ calculations. On the other hand, the W1 and W2 procedures which use rigorous basis set extrapolations from reasonably large Gaussian basis sets result in near-zero absolute deviations and percentage errors ≤ 5%. We note that the W1 procedure has a slight edge over the CCSD(T)/A'VTZ + MP2/A'V{T,Q}Z – MP2/A'VTZ procedure at a similar computational cost.

**Conclusions**

The benzene•••ethene and benzene•••benzene dimers are the most fundamental systems involving π-π stacking interactions. A number of high-level ab initio investigations calculated the binding energies of these dimers, however, these studies relied on MP2-based basis set correction schemes and/or truncated optimized virtual orbital spaces. In the present work we obtain the CCSD(T) binding energies at the infinite one-particle basis set limit without relying on such approximations via a W3-type composite approach. In particular, we extrapolate the SCF, CCSD, and (T) components using basis sets of up to A'V6Z, A'V5Z, and A'VQZ quality, respectively. In addition, we consider post-CCSD(T) contributions up to CCSDT(Q), inner-shell, scalar-relativistic, and Born–Oppenheimer corrections. Our best valence CCSD(T)/CBS binding energies are 1.244 (benzene•••ethene) and 2.580 (benzene•••benzene) kcal mol$^{-1}$. Our best all-electron, relativistic, DBOC-inclusive CCSDT(Q) binding energies are $\Delta E_{e,\text{all,rel}}$ = 1.234 (benzene•••ethene) and 2.550 (benzene•••benzene). Converting these $\Delta E_{e,\text{all,rel}}$ values to binding energies at 0 K ($\Delta H_0$) using scaled harmonic DSD-PBEP86-D3BJ/Def2-QZVPPD ZPVEs results in $\Delta H_0$ = 0.949 (benzene•••ethene) and 2.310 (benzene•••benzene). The latter value is in excellent agreement with the experimental value of $\Delta H_0$ = 2.4 ± 0.4 kcal mol$^{-1}$, but suggests that the experimental value of $\Delta H_0$ = 1.6 ± 0.2 kcal mol$^{-1}$ should be significantly revised upwards.

Our best MP2 and CCSD limits agree to within expected uncertainties with values obtained from explicitly correlated MP2-F12 and CCSD(F12*) calculations.

There is an effective cancellation between the T–(T) and (Q) components (i.e., they have opposite signs and are of similar magnitude). Overall, post-CCSD(T) contributions reduce the binding energies by 0.028 (benzene•••ethene) and 0.058 (benzene•••benzene) kcal mol$^{-1}$, thus they cannot be neglected if sub-chemical accuracy is sought (i.e., errors below 0.1 kcal mol$^{-1}$). CCSD(T)/aug-cc-pwCVTZ core-valence corrections increase the binding energies by 0.018 (benzene•••ethene) and 0.027 (benzene•••benzene) kcal mol$^{-1}$. Scalar-relativistic and diagonal Born–Oppenheimer corrections are negligibly small.

We use our best CCSDT(Q) binding energies to evaluate the performance of a range of standard and composite ab initio procedures. With the exception of SCS-MP2, all the MP2-based procedures result in relatively large percentage errors of over 10% close to the CBS limit. The composite G4- and CBS-type procedures also result in relatively large percentage errors of over 50% in most cases. The CCSD(T)/A'VTZ + MP2/A'V{T,Q}Z – MP2/A'VTZ additivity scheme as well as the W1 and W2 composite ab initio methods result in excellent performance with percentage errors < 5%.

## Supplementary data

Optimized DSD-PBEP86-D3BJ/Def2-QZVPPD geometries for all structures (Tables S1 and S2); basis set convergence of the MP$n$-based procedures presented in Table 7 (Table S3); Key absolute energies (Table S4); and full references for quantum chemical software.

## Acknowledgments

This research was undertaken with the assistance of resources from the National Computational Infrastructure (NCI), which is supported by the Australian Government. We also acknowledge the system administration support provided by the Faculty of Science at the University of Western Australia to the Linux cluster of the Karton group. We would like to thank the reviewers of the manuscript for their valuable comments and suggestions.

## Funding Information

Australian Research Council (ARC) Future Fellowship (Project No. FT170100373); Israel Science Foundation (grants 1358/15 and 1969/20).

## Data Availability Statement

The data that supports the findings of this study are available within the article [and its supplementary material]. Any additional data are available from the corresponding author upon reasonable request.

[54] We note that at the DSD-PBEP86-D3BJ/Def2-QZVPPD level of theory we obtain a small imaginary frequency of 3.9 cm$^{-1}$ for the TT dimer which is attributed to numerical instability. At the DSD-PBEP86-D3BJ/Def2-TZVPPD level of theory we obtain all real frequencies for the TT dimer.